\documentclass[%
 aip,
rsi,%
 amsmath,amssymb,
 preprint,%
]{revtex4-1}
\pdfoutput=1
\usepackage{graphicx}
\usepackage{dcolumn}
\usepackage{bm}
\usepackage{amsmath}
\usepackage{color}
\usepackage{soul}
\usepackage{bbm} 
\usepackage{amssymb} 
\usepackage{gensymb} 
\usepackage{epstopdf}

\newcommand{\ve}[1]{\boldsymbol{#1}}  
\newcommand{\pd}[2]{\frac{\partial #1}{\partial #2}} 
\newcommand{\fd}[2]{\frac{\delta #1}{\delta #2}} 

\graphicspath{{./images/}}

\begin{document}

\title{A microscopic field theoretical approach for binary mixtures of active and passive particles}

\author{Francesco Alaimo}
\affiliation{Institute of Scientific Computing, Technische Universit\"at Dresden, 01062 Dresden, Germany}
\affiliation{Dresden Center for Computational Materials Science (DCMS), 01062 Dresden, Germany}

\author{Axel Voigt}
\email[Corresponding author: ]{axel.voigt@tu-dresden.de.}

\affiliation{Institute of Scientific Computing, Technische Universit\"at Dresden, 01062 Dresden, Germany}
\affiliation{Dresden Center for Computational Materials Science (DCMS), 01062 Dresden, Germany}
\affiliation{Center of Systems Biology Dresden (CSBD), Pfotenhauerstr. 108, 01307 Dresden, Germany}


\begin{abstract}
We consider a phase field crystal modeling approach for binary mixtures of interacting active and passive particles. 
The approach allows to describe generic properties for such systems within a continuum model. 
We validate the approach by reproducing experimental results, as well as results obtained with agent-based simulations, 
for the whole spectrum from highly dilute suspensions of passive particles to interacting active particles in a dense background of passive particles.    
\end{abstract}

\maketitle
\section{Introduction}  

  Active systems have been in the focus of intense research for the last decade because they provide deep insights into the self-organization of systems that are 
  intrinsically in a non-equilibrium state such as living matter. Even more interesting are mixtures of active and passive particles. Situations of active particles in 
  crowed environments or passive particles in an active bath resemble the situation in living matter more realistically and might even shed light on the active dynamic 
  processes within a cell \cite{Dasetal_PRL_2016}. Observed phenomena in mixtures of active and passive particles are e.g. activity-induced phase-separation 
  \cite{Stenhammar2015}, the formation of large defect-free crystalline domains \cite{Kummel2015}, propagating interfaces \cite{Wysocki2016} but also 
  a transition from diffusive to subdiffusive dynamics \cite{Zeitz2017,morin2017_a} and suppressed collective motion \cite{morin2017_b}. To understand this wide span 
  of phenomena is crucial to almost all applications of active systems.
  
  Different ways exist to describe phenomena in active systems theoretically. Typical approaches for active systems consider either the microscopic scale, taking 
  the interactions between the particles into account, or the macroscopic scale, focusing on the emerging phenomena. For reviews on both theoretical descriptions 
  see e.g. \cite{Marchetti2013, Ramaswany2010}. Examples for extensions towards mixtures of active and passive particles are summarized in \cite{Bechinger2016}
  and range from active particles in confined domains \cite{briand2017, Lushi2013, Wioland2013}, 
  through active particles moving between fixed or moving obstacles \cite{Reichhardt2017, Reichhardt2018, Kaiser2012, Kaiser2013, Zeitz2017},
  to binary mixtures of interacting active and passive particles \cite{McCandlish2012, Kummel2015, Stenhammar2015, Wysocki2016}. All these studies are 
  examples for models on the microscopic scale. In \cite{Menzel2013,Alaimo2016} a continuum modeling approach was introduced for active systems which combines aspects 
  from the microscopic and the macroscopic scales. The goal of the paper is to extend this approach to mixtures of active and passive particles which will allow to 
  describe generic properties of such systems. We use the model to study 
  the effect of a few active particles in passive systems (active doping) \cite{Kummel2015, Ni2014, VanderMeer2016} and 
  how passive particles perturb collective migration in an active bath \cite{Wu2000, Valeriani2011, Hinz2014, Zeitz2017}.
  For the first case we observe enhanced crystallization in the passive system, in quantitative agreement with the results of \cite{Kummel2015}.
  For the active bath case we investigate how collective migration is affected by a
  disordered environment. For the special case of immobile passive particles these results are in agreement with \cite{morin2017_a, morin2017_b}. However, for mobile
  passive particles new phenomena and patterns emerge, which ask for experimental validation. Also the intermediate regime of similar fractions of active and passive 
  particles is rich in complex phenomena but much harder to quantify.

\section{The Model}
  The starting point for the derivation of the model is the microscopic field theoretical model for active particles introduced in \cite{Alaimo2016}. It has been validated against 
  known results obtained with minimal agent-based models and proven to be applicable for large scale computations. The model reads in scaled units
    \begin{equation}\label{EQ::Menzel}
  \begin{array}{l}
    \partial_t \psi = M_0 \Delta \frac{\delta \mathcal{F}_{\rm vpfc}}{\delta \psi} - v_0 \triangledown \cdot (\psi \mathbf{P}) \\
    \partial_t \mathbf{P} = \Delta \left(\alpha_2 \mathbf{P} + C_2 \mathbf{P}^3 \right) - \left( \alpha_4 \mathbf{P} + C_4 \mathbf{P}^3 \right) - v_0 \mathbf{\triangledown} \psi - \beta \mathbf{P} \mathbbm{1}_{\psi_A \leq 0 } 
  \end{array}
  \end{equation}
 for a one-particle density field $\psi(\ve{r},t)$, which is defined with respect to a reference density $\bar{\psi}$, and the polar order parameter $\mathbf{P}(\ve{r},t)$, which is related to a 
 coarse-grained velocity field with a typical magnitude $v_0$ of the self-propulsion velocity. $\mathbf{P}$ is a local quantity that
 is different from zero only within the peaks of the density field $\psi$, which is ensured by $\beta > 0$ typically larger than the other terms entering the $\mathbf{P}$ equation. $M_0$ is the mobility, $\alpha_2$ and $\alpha_4$ are two parameters related to relaxation 
 and  orientation of the polarization field and $C_2$ and $C_4$ are parameters which govern the local orientational ordering. For simplicity we will restrict ourselves to the case 
 $C_2 = C_4 = 0$, which only allows gradients in the density field $\psi$ to induce local polar order. The energy functional 
 $\mathcal{F}_{\rm vpfc} = \mathcal{F}_{\rm pfc} + \mathcal{F}_{\rm v} $ consists of a Swift-Hohenberg energy \cite{SwiftHohenberg1977}
  \begin{equation}\label{EQ::PFC_energy}
    \mathcal{F}_{\rm pfc} = \int \left[ \frac{1}{4}\psi^4 + \frac{1}{2} \psi (q + (1 + \Delta)^2)\psi \right] d \mathbf{r},
  \end{equation}
  with a parameter $q$ related to temperature and a penalization term 
  \begin{equation}\label{EQ::V_energy}
    \mathcal{F}_{\rm v} = \int H(|\psi|^3 - \psi^3) d \mathbf{r},
  \end{equation}
  with $H \simeq 1500$ to constrain the one-particle density field $\psi$ to positive values. The penalization term $\mathcal{F}_{\rm v}$ is the essential modification which allows to model individual particles 
  \cite{Chan2009, Berry2011, Robbins2012, Praetorius2015}. Without this additional term the model can be related to models for active crystals \cite{Menzel2013, Menzel2014}. If, in addition, we neglect the coupling with the polar order parameter $\mathbf{P}$ we obtain the classical phase field crystal (PFC) model  introduced in \cite{Elder2002, Elder2004} to model 
  elasticity in crystalline materials. For a detailed derivation of (\ref{EQ::PFC_energy}) and its relation to classical density functional theory we refer to \cite{Elder2007, vanTeeffelen2009}.
  If the coupling with $\mathbf{P}$ is neglected but the penalization term (\ref{EQ::V_energy}) considered, the model is known as the vacancy PFC (VPFC) model  \cite{Chan2009}.

  Various ways have been introduced to extend the classical PFC model towards a second species, thus modeling binary mixtures \cite{Elder2007,Robbins2012}. We adopt one 
  of these approaches for the VPFC model by considering energies for species $A$ and $B$ with
   \begin{equation}\label{EQ::BVPFC_energy}
    \mathcal{F}(\psi_A, \psi_B) = \mathcal{F}^A_{\rm vpfc}(\psi_A) + \mathcal{F}^B_{\rm vpfc}(\psi_B) + \mathcal{F}_{\rm int}^{AB}(\psi_A, \psi_B)
  \end{equation}
  where $\mathcal{F}^i_{\rm vpfc}, i = A,B$ as before and
  \begin{equation}\label{EQ::binary_interaction}
  \mathcal{F}_{\rm int}^{AB}(\psi_A, \psi_B) = \frac{a}{2} \psi_A^2 \psi_B^2 
  \end{equation}
  an interaction energy with $a > 0$. 
  
  In principle both species appearing in (\ref{EQ::BVPFC_energy}) could be made active. 
  Our aim is however to simulate binary mixtures of interacting active and passive particles. With
  this in mind we couple only species A to the polar order parameter $\mathbf{P}$. We further assume
  $\mathcal{F}^A_{\rm vpfc} = \mathcal{F}^B_{\rm vpfc} = \mathcal{F}_{\rm vpfc}$ and thus e.g. equal size of the active and passive particles.
  The resulting dynamical equations are
  \begin{equation}\label{EQ::dynamics} 
    \begin{array}{ll}
    \pd{\psi_A}{t} = M_0^A \Delta \left( \fd{\mathcal{F}(\psi_A)}{\psi_A} +  a \psi_A \psi_B^2 \right) 
			  - v_0 \nabla \cdot (\psi_A \mathbf{P}) \\

    \partial_t \mathbf{P} = \alpha_2 \Delta \mathbf{P} - \alpha_4 \mathbf{P}  
    - v_0 \nabla \psi_A - \beta \mathbf{P} \mathbbm{1}_{\psi_A \leq 0 } \\

    \pd{\psi_B}{t} = M_0^B \Delta \left( \fd{\mathcal{F}(\psi_B)}{\psi_B} + a \psi_A^2 \psi_B \right)
    \end{array}
  \end{equation} 
 which define a microscopic field theoretical approach for binary mixtures of interacting active and passive particles. An extension to more than two species, species with different size 
 and interaction potential and active species with different self-propulsion velocities is obvious.

\section{Results} 
  We solve equations (\ref{EQ::dynamics}) in two dimensions using a parallel finite element approach \cite{Backofen2007}. We adopt a block-Jacobi preconditioner \cite{Praetorius2015_SIAM} 
  that allows us to use a direct solver locally. This is implemented in AMDiS \cite{Vey2007, Witkowski2015}. 
  The computational domain is a square of size $L = 200$ with periodic boundary conditions. 
  The initial condition for $\psi_A$ and $\psi_B$ is calculated using a one-mode approximation with lattice distance $d = 4 \pi/ \sqrt{3} $ 
  for each particle \cite{Praetorius2015}, with the centers placed randomly according to a packing algorithm \cite{Skoge2006}. 
  The $\mathbf{P}$ field is set to zero initially.

  Each maxima in the one-particle density fields $\psi_A$ and $\psi_B$ is interpreted as an active or passive particle, respectively. The diameter of the particle is defined by the lattice 
  distance $d$. We track the particle positions $\mathbf{x}^i_{A,B}(t)$ and use this information to compute 
  the particle velocities $\mathbf{v}^i_{A,B}(t)$ as the discrete time derivative of two successive maxima. We define the total particle density $\phi = N \sigma/L^2$, 
  with $N$ the total number of particles $N = N_A + N_B$ and $N_{A,B}$ the number of $A$ and $B$ particles, respectively. The parameter $\sigma = \pi (d/2)^2$ is the area 
  occupied by a single
  particle. The relative density $\phi_A = N_A/N$ corresponds to the fraction of active particles present in the system.
  For low relative densities ($\phi_A < 0.2$) we are in the regime of active doping, and analyse how a passive system is influenced by the presence of a few active particles.
  For high relative densities ($\phi_A > 0.7$) we are in the regime of an active bath and study how a few passive particles affect an active system. 

  We fix the following parameters $(a, v_0, \alpha_2, \alpha_4, \beta, H, q) = (200, 1.5, 0.2, 0.1, 2, 1500, -0.9)$, unless otherwise specified in the figure captions.

  \subsection{Active doping: how active particles enhance crystallization}
  
    \begin{figure}
      \centering
      \includegraphics{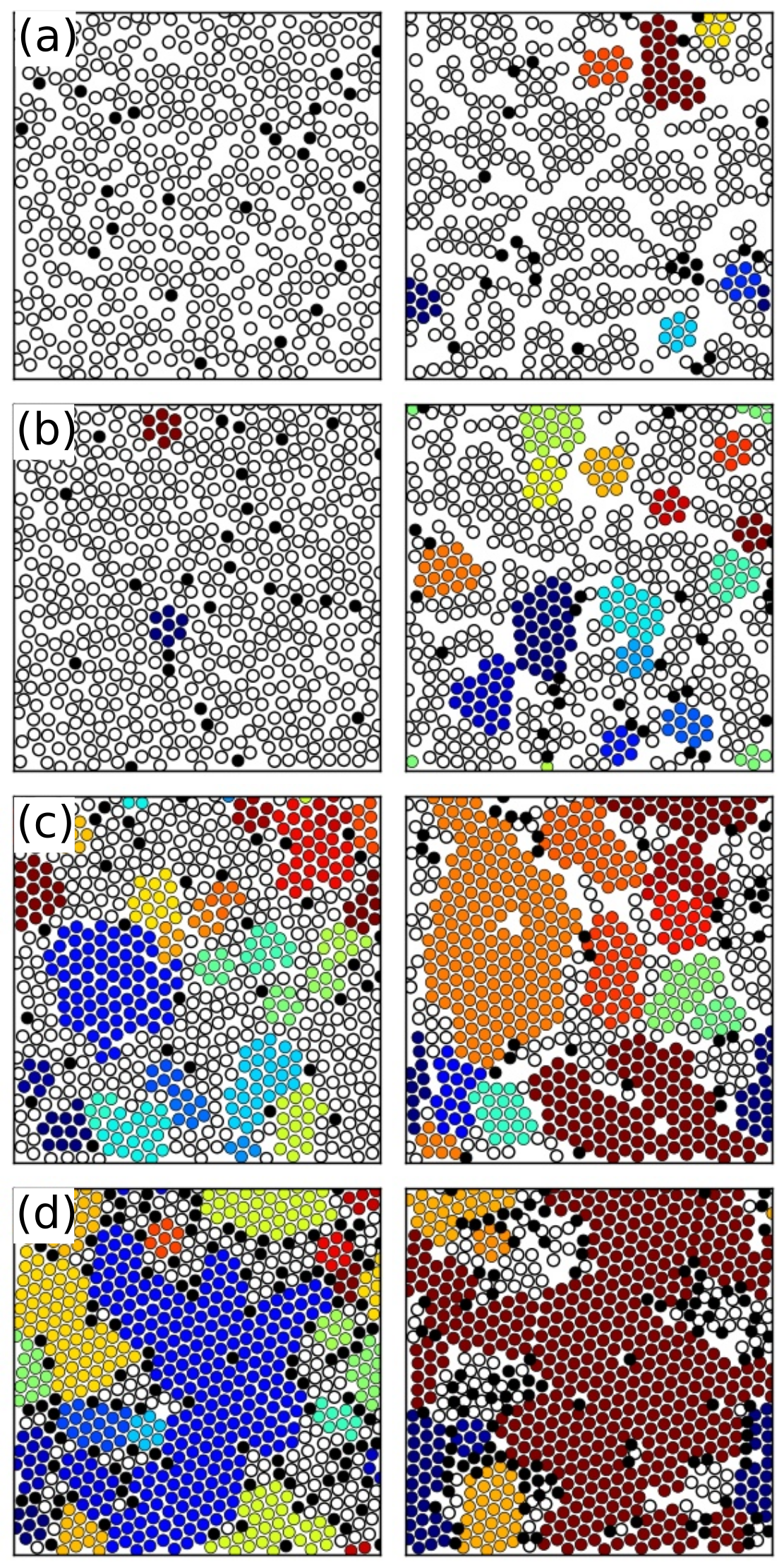}\\
      \caption{Snapshots showing passive clusters for different relative and absolute
	densities at time $ t = 0$ (first column) and at time $t = 1000$ (second column).
	Particles with the same color belong to the same cluster,
	white disks represents passive particles not belonging to any cluster and black disks are active particles.
	(a) $\phi = 0.5, \phi_A = 0.05$, (b) $\phi = 0.6, \phi_A = 0.05$
	(c) $\phi = 0.7, \phi_A = 0.05$, (d) $\phi = 0.8, \phi_A = 0.15$.
	Other parameters are $M_0^A=M_0^B = 50$.}
      \label{figure1}
    \end{figure}

    It has been shown by particle simulations \cite{Ni2014, VanderMeer2016} and experimentally \cite{Kummel2015} that the crystalline structure of passive 
    particles is altered by the presence of active agents. More precisely active particles generate density variation in the passive system and promote crystallization, 
    leading to the formation of passive clusters. To analyze these phenomena with our microscopic field theoretical approach we need to identify if a particle belongs to a cluster. 
    We follow the definition of \cite{Kummel2015} where two criteria have to be fulfilled. The nearest neighbor distances less than $3/2 d$ and the coordination
    number is $6$.
    
    \begin{figure}
      \centering
      \includegraphics{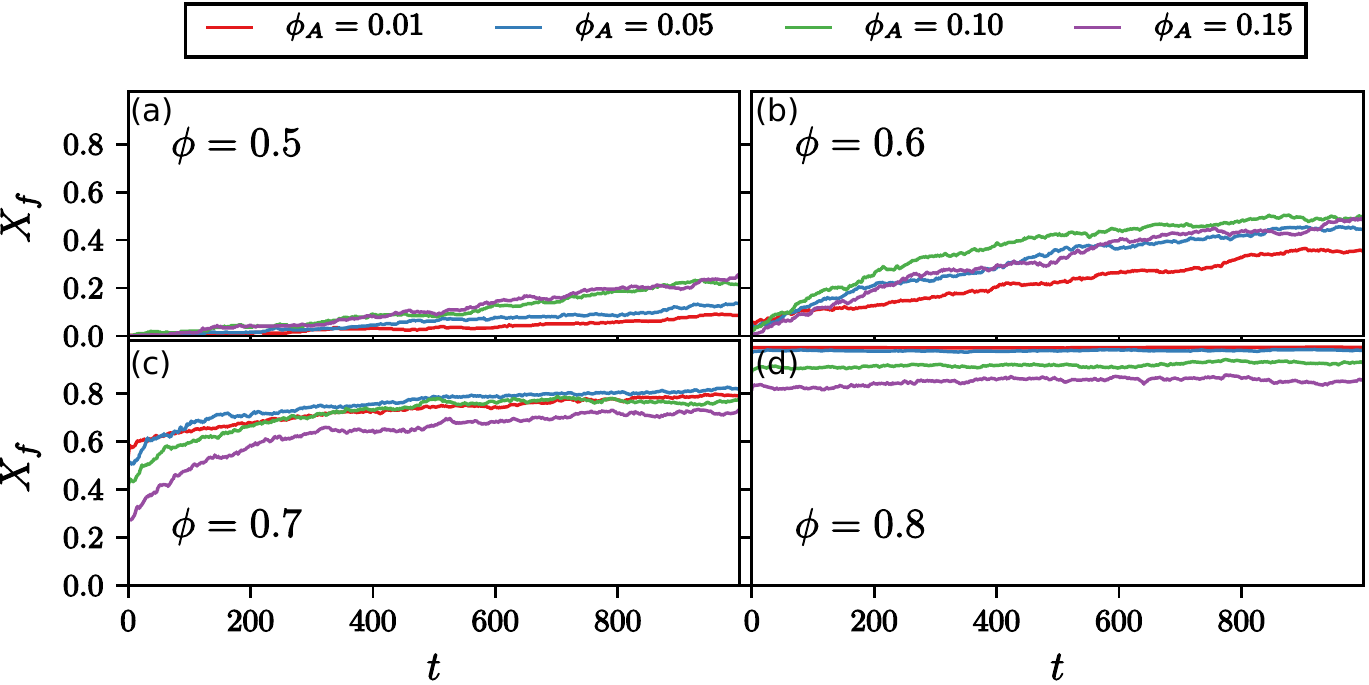}\\
      \caption{Percentage of passive particles belonging to a cluster $X_f$ as a function of time for different relative and absolute densities $\phi_A$ and $\phi$.
	We observe that for $\phi = 0.5$ and $\phi = 0.6$ (top row), increasing
	the number of active particles lead to an increase of $X_f$, whereas the opposite
	is true for $\phi = 0.7$ and $\phi = 0.8$ (bottom row).
	Other parameters are $M_0^A=M_0^B = 50$.
	Each curve has been obtained as the average of five different simulations started with
	different initial conditions.}
      \label{figure2}
    \end{figure}

    \begin{figure}
      \centering
      \includegraphics{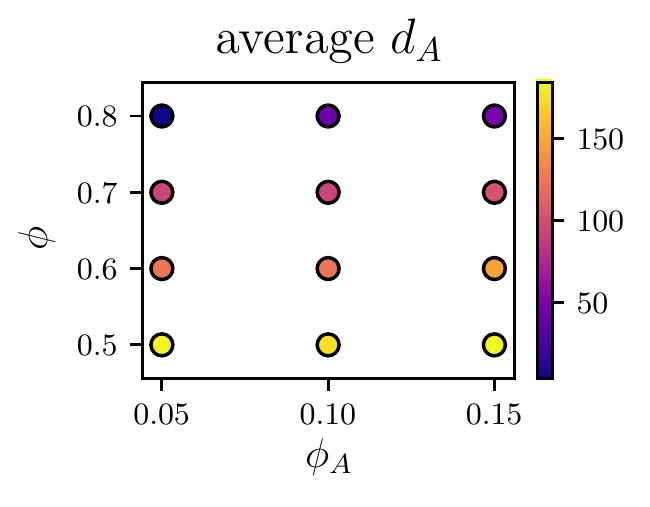}\\
      \caption{Average of the maximum displacement $d_A$ of active particles moving in a binary mixture for different values of $\phi$ and $\phi_A$.
	Active particles travel a longer distance when the passive particles have not crystallized, until the extreme case of $\phi = 0.8$, $\phi_A = 0.05$
	where $d_A$ is so low that active particles are basically trapped (see also supplementary video).
	Other parameters are $M_0^A=M_0^B = 50$. 
	Each point has been obtained as the average of five different simulations started with
	different initial conditions.}
      \label{figure3}
    \end{figure}

    Figure \ref{figure1} shows snapshots with passive clusters for different relative and absolute densities, $\phi_A$ and $\phi$, respectively. The time evolution of the percentage of 
    passive particles which belong to a cluster $X_f$ is shown in figure \ref{figure2}.
    For dilute systems ($\phi = 0.5$, figure \ref{figure1}(a)) $X_f$ slowly increases with time. Increasing the relative density $\phi_A$ leads to larger values $X_f$. However, 
    it remains relatively low, rarely exceeding $20 \%$, for the considered time ($t= 1000$).
    Increasing the density ($\phi = 0.6$, figure \ref{figure1}(b)) the system changes from a state where no clusters are present ($t = 0$)
    to a state where up to $50\%$ of the passive particles are found in clusters. A maximum $X_f$ 
    is observed for $\phi_A = 0.1$, where $X_f$ saturates at $t = 1000$. Further increasing the number of active particles leads to a reduction of $X_f$. 
    Adding more and more active particles to systems with already existing crystalline clusters introduces disorder. A phenomena already observed in \cite{Kummel2015}. 
    By further increasing the density ($\phi = 0.7$, figure \ref{figure1}(c)) some clusters are already present for the random initial configuration at $t = 0$, due to spontaneously 
    crystallization. Active particles can be inside these regions, thus disturbing their symmetry. This
    explains why the system behaves in the opposite way as for the dilute case, with $X_f$ decreasing as 
    the fraction of active particles $\phi_A$ increases. 
    Finally for $\phi = 0.8$ the initial configuration is already almost completely crystallized ($X_f \simeq 1$ for $t = 0$, figure \ref{figure1}(d)). 
    Adding active particles partially destroys the crystalline structure (figure \ref{figure2}) and $X_f$ decreases for increasing $\phi_A$. We thus observe both phenomena, 
    enhanced crystallization in dilute systems and suppressed crystallization in dense systems. 

    A final observation concerns how the dynamics of the active particles is affected by the presence of passive ones.
    In figure \ref{figure3} the maximum displacement $d_A$ for active particles (averaged over all the particles) is shown as a function 
    of the absolute and relative densities $\phi$ and $\phi_A$. No data is shown for $\phi_A = 0.01$, as the  number of active particles
    is too small for meaningful averages. We observe a clear correlation between $d_A$ and 
    the crystallization in the system: the higher $X_f$, the smaller is the maximum displacement of active particles until,
    for the extreme case of $\phi = 0.8$ and $\phi_A= 0.05$ active particles are trapped inside a big cluster and have a very small maximum displacement
    (see also supplementary video).
  
  \subsection{Active bath: how passive particles can suppress collective migration}
    Inelastic collisions in systems which are composed solely of active particles can lead to collective motion. This has been shown by
    particle based models, e.g. \cite{Grossman2008}, microscopic field theoretical models \cite{Alaimo2016} and phase field models \cite{Lober2015, Marth2016}. 
    In all these models the state of collective motion is characterized by the translational 
    order parameter $\phi_T = 1/N_A \left | \sum_{i = 1}^{N_A} \hat{\mathbf{v}}^i_A(t) \right |$ being close to one, with $\hat{\mathbf{v}}^i_A(t)$ the unit velocity vector
    for the active particle $i$ at time $t$. We here analyze the stability of the state of collective motion, if passive particles are
    introduced in the system. How do the relative and absolute densities and the mobility of the passive particles affect this state?

    \begin{figure}
      \centering
      \includegraphics{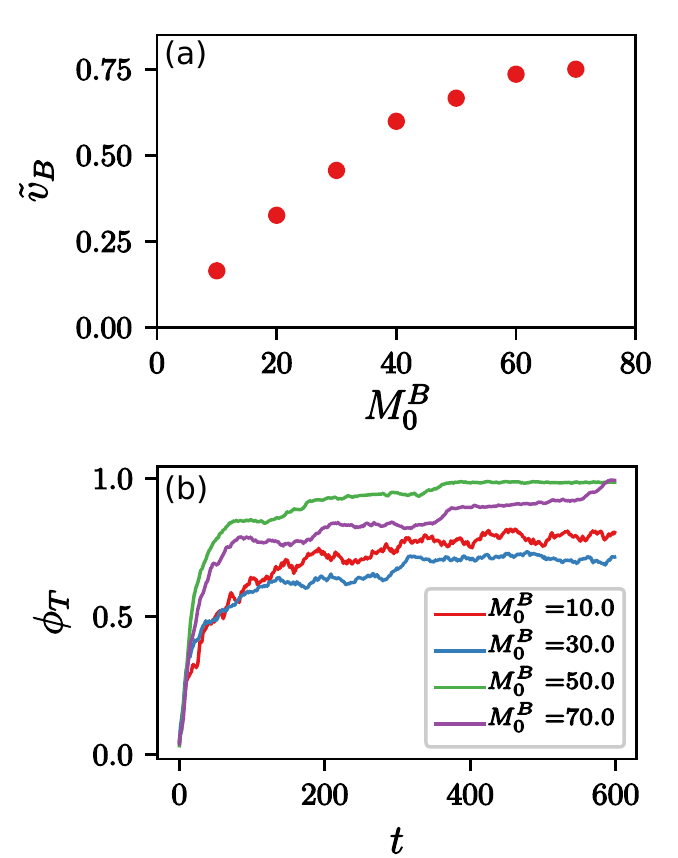}\\
      \caption{(a) Average velocity $\tilde{v}_B$ of passive particles as a function of their mobility in an active bath with $\phi = \phi_A = 0.9$.
	$\tilde{v}_B$ increases almost linearly for small $M_0^B$ until it starts to saturate at around $M_0^B = 70$. 
	(b) Translational order parameter $\phi_T$ as a function of time for different mobility $M_0^B$. For small values of $M_0^B$
	there is no collective migration, for intermediate values this state is reached quite fast, 
	whereas for high mobility the transient phase to reach collective migration increases. 
	However simulations are not so numerically stable for small and intermediate values of the mobility and this is why we choose $M_0^B = 70$
	for the analysis in figure \ref{figure5}.
	$M_0^A = 100$ for both cases.
	The data have been obtained as the average of ten different simulations started with
	different initial conditions.}
      \label{figure4}
    \end{figure}

    To consider a dense system we fix $\phi = 0.9$. We further set $\phi_A = 0.9$ and vary the mobility of the few passive particles $M_0^B$. For low mobilities they act as fixed objects
    and the results can be compared with experimental studies for active colloids in disordered environments \cite{morin2017_b}, which show a suppression of collective motion. Also in  
    our simulations the active system does not reach a state of collective motion, as shown from the time series of $\phi_T$ (figure \ref{figure4}(b)). However, the situation changes if the
    passive particles are mobile. Figure \ref{figure4}(a) shows the average velocity $\tilde{\mathbf{v}}_B$ of the passive particles as a 
    function of their mobility. Increasing $M_0^B$, the average passive particles velocity $\tilde{\mathbf{v}}_B$ also increases, meaning that passive particles are
    transported from the active ones. For $M_0^B > 30$ a state of collective migration is reached (figure \ref{figure4}(b)), even though 
    the time required to reach it is larger than in the homogeneous case $\phi_A = 1$ (no passive particles present).

    \begin{figure}
      \centering
      \includegraphics{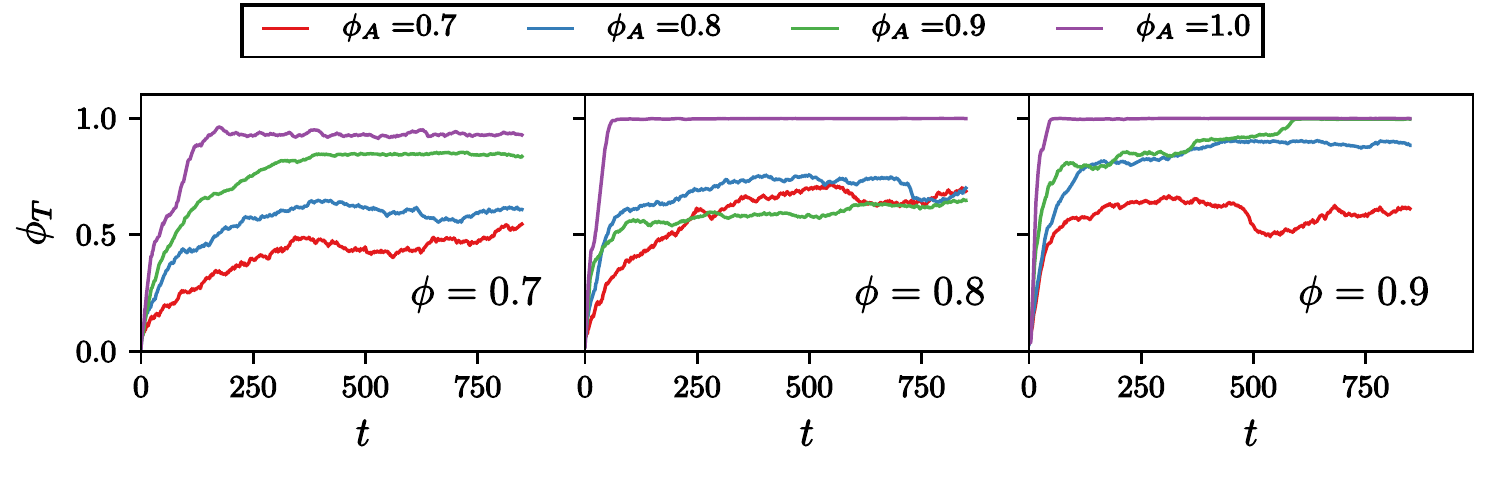}\\
      \caption{Translational order parameter $\phi_T$ as a function of time for different values of $\phi$ and $\phi_A$. 
	The purple curve corresponds to the case $\phi_A = 1$, i.e. no passive particles present. We see that in all other cases the
	state of collective migration is reached later (longer transient phase) or not reached at all, especially for lower $\phi_A$ (red curves).
	Other parameters are $M_0^A=100$ and $M_0^B = 70$. The data have been obtained as the average of ten different simulations started with
	different initial conditions.}
      \label{figure5}
    \end{figure}

\begin{figure}
      \centering
      \includegraphics{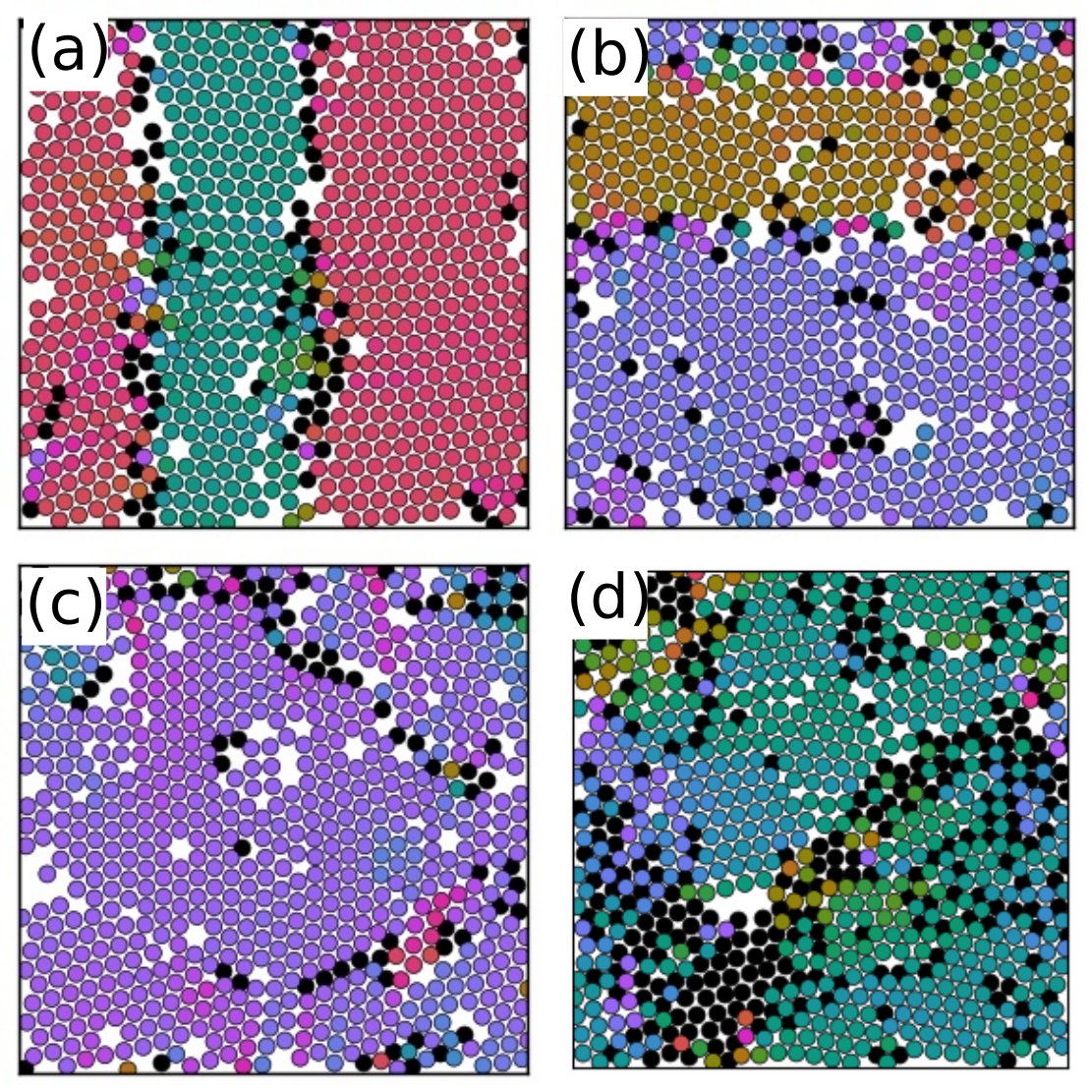}\\
      \caption{ The color code corresponds to the single particle velocity.
	(a) Snapshot of a state of local flocking, with two macro regions of active particles having exactly opposite orientation.
	This state can last for a long time thanks to the presence of passive particles at the boundary between the two regions (see also
	supplementary video).
	(b) Another state of local flocking. Here, less passive particles are accumulated at the boundary between the moving active regions. 
	The situation is less stable and transforms in a situation of global collective motion (c), in which passive particles form chains, which persist
	over longer periods of time. (d) Passive particles forming clusters in an active bath. (a) - (c) regime $\phi = 0.8$, $\phi_A = 0.9$, (d) regime $\phi = 0.9$, $\phi_A = 0.7$.}
      \label{figure6}
    \end{figure}

    We now fix the mobility $M_0^B = 70$ and vary $\phi$ and $\phi_A$.
    We reduce $\phi$ down to $0.7$, a limit for which a state of collective migration would still be 
    reached in an homogeneous active system ($\phi_A=1$), as seen from the purple lines in figure \ref{figure5}.
    For $\phi = 0.9$ a state of collective migration is reached for $\phi_A = 0.9$ but with a longer transient phase than for the homogeneous case 
    (green line in figure \ref{figure5}(c)). For $\phi_A = 0.8$ we already see a small 
    perturbation from the unit value for $\phi_T$ and for $\phi_A = 0.7$ collective migration is no longer reached.  
    We here observe the accumulation of passive particles in certain regions, see also figure \ref{figure6}(d). This hinders
    the active particles from following a straight trajectory and thus the formation of collective migration.  
    Things change by reducing the total density to $\phi = 0.8$. The state of collective migration is not reached, independent of the value of $\phi_A$ (figure \ref{figure5}(b)). 
    However, for high relative density $\phi_A = 0.9$, green curve in figure \ref{figure5}(b) a new state is formed, where the
    translational order parameter $\phi_T$ is at least locally close to one. This new state is discussed below and can be seen
    in the snapshots in figures \ref{figure6}(a) and (b).       
    For $\phi = 0.7$ (figure \ref{figure5}(a)) a decrease in $\phi_A$ leads to a decrease of $\phi_T$. In this situation there is enough empty space in the system to allow
    active particles to change their trajectories when interacting with passive ones. This causes a perturbation that
    gets bigger as the number of passive particles increase, leading to a decrease of $\phi_T$.
    
    A more detailed investigation of the intermediated regime with $\phi = 0.8$ and $\phi_A = 0.9$ is shown in figure \ref{figure6}(a) and (b), showing an intermediate state 
    with two regions of active particles moving in opposite direction. The regions are separated by passive particles. This separation prevents the alignment of the collectively 
    migrating domains. This state can be seen as a local flocking state. It is more stable in figure \ref{figure6}(a), persisting for the whole simulation time, and less stable in 
    figure \ref{figure6}(b), where the alignment of passive particles will be destroyed after a while and a transition to collective migration follows (see supplementary video).
    However, even if this collective migration state is reached the passive particles are not randomly distributed. As see in figure \ref{figure6}(c) they form chains, which persist over
    longer periods of time and are transported by the active particles. If the number of passive particles is increased $\phi_A = 0.7$ a clustering of passive particles within the active bath
    can be observed, see figure \ref{figure6}(d). These new states and patterns are characteristic for binary mixtures and should be explored further, both numerically and experimentally.

\section{Conclusions}
  In summary, our microscopic field theoretical approach for binary mixtures of interacting active and passive particles has been used to investigate a wide spectrum from dilute systems
  $\phi < 0.7$ to dense systems $\phi > 0.7$ with a relatively low fraction of active particles $\phi_A < 0.2$ (active doping) and a relatively high fraction $\phi_A > 0.7$ (active bath), 
  respectively. We 
  have demonstrated with one and the same model a variety of known phenomena, such as enhanced crystallization via active doping \cite{Kummel2015, Ni2014} and suppressed 
  crystallization in dense systems \cite{Kummel2015}. We also analyzed the limits of collective migration, which for the special case of immobile passive particles qualitatively 
  reproduce the results in \cite{morin2017_b}. Within the experiments in \cite{Kummel2015} and in our simulations the suppression of collective migration sensitively depends on the 
  fraction of immobile passive particles. Within the experimentally less 
  explored state of mobile passive particles we found new phenomena. For fractions of passive particles, for which collective migration is suppressed if the passive particles are immobile,
  collective motion is still possible if the mobility of these particles is large enough. 
  But there are also intermediate regime, characterized by local flocking states, where regions of active particles 
  are separated by boundary layers of passive ones. We further found chains of passive particles and clusters which persist for a relatively long time. 
  A rigorous classification of these states remains open and should be addressed with experimental investigations. 
  
  As already pointed out, the proposed microscopic field theoretical model can easily be modified to consider more than two species, species with different size and interaction potential 
  and active species with different self-propulsion velocities, which makes the approach a generic tool to study active systems in complex environments. Also hydrodynamic interactions have 
  already be considered together with a (passive) phase field crystal model \cite{Praetorius2015, Heinonen2016} and they
  could also be included in our model.

\begin{acknowledgments}
  This work is funded by the European Union (ERDF) and the Free State of Saxony via the ESF project 100231947 
  (Young Investigators Group Computer Simulation for Materials Design - CoSiMa). 
  We used computing resources provided by JSC within project HDR06.
\end{acknowledgments}

\clearpage

\bibliography{bibliography}

\begin{thebibliography}{44}%
\makeatletter
\providecommand \@ifxundefined [1]{%
 \@ifx{#1\undefined}
}%
\providecommand \@ifnum [1]{%
 \ifnum #1\expandafter \@firstoftwo
 \else \expandafter \@secondoftwo
 \fi
}%
\providecommand \@ifx [1]{%
 \ifx #1\expandafter \@firstoftwo
 \else \expandafter \@secondoftwo
 \fi
}%
\providecommand \natexlab [1]{#1}%
\providecommand \enquote  [1]{``#1''}%
\providecommand \bibnamefont  [1]{#1}%
\providecommand \bibfnamefont [1]{#1}%
\providecommand \citenamefont [1]{#1}%
\providecommand \href@noop [0]{\@secondoftwo}%
\providecommand \href [0]{\begingroup \@sanitize@url \@href}%
\providecommand \@href[1]{\@@startlink{#1}\@@href}%
\providecommand \@@href[1]{\endgroup#1\@@endlink}%
\providecommand \@sanitize@url [0]{\catcode `\\12\catcode `\$12\catcode
  `\&12\catcode `\#12\catcode `\^12\catcode `\_12\catcode `\%12\relax}%
\providecommand \@@startlink[1]{}%
\providecommand \@@endlink[0]{}%
\providecommand \url  [0]{\begingroup\@sanitize@url \@url }%
\providecommand \@url [1]{\endgroup\@href {#1}{\urlprefix }}%
\providecommand \urlprefix  [0]{URL }%
\providecommand \Eprint [0]{\href }%
\providecommand \doibase [0]{http://dx.doi.org/}%
\providecommand \selectlanguage [0]{\@gobble}%
\providecommand \bibinfo  [0]{\@secondoftwo}%
\providecommand \bibfield  [0]{\@secondoftwo}%
\providecommand \translation [1]{[#1]}%
\providecommand \BibitemOpen [0]{}%
\providecommand \bibitemStop [0]{}%
\providecommand \bibitemNoStop [0]{.\EOS\space}%
\providecommand \EOS [0]{\spacefactor3000\relax}%
\providecommand \BibitemShut  [1]{\csname bibitem#1\endcsname}%
\let\auto@bib@innerbib\@empty
\bibitem [{\citenamefont {Das}, \citenamefont {Polley},\ and\ \citenamefont
  {Rao}(2016)}]{Dasetal_PRL_2016}%
  \BibitemOpen
  \bibfield  {author} {\bibinfo {author} {\bibfnamefont {A.}~\bibnamefont
  {Das}}, \bibinfo {author} {\bibfnamefont {A.}~\bibnamefont {Polley}}, \ and\
  \bibinfo {author} {\bibfnamefont {M.}~\bibnamefont {Rao}},\ }\href {\doibase
  10.1103/PhysRevLett.116.068306} {\bibfield  {journal} {\bibinfo  {journal}
  {Phys. Rev. Lett.}\ }\textbf {\bibinfo {volume} {114}},\ \bibinfo {pages}
  {068306} (\bibinfo {year} {2016})}\BibitemShut {NoStop}%
\bibitem [{\citenamefont {Stenhammar}\ \emph {et~al.}(2015)\citenamefont
  {Stenhammar}, \citenamefont {Wittkowski}, \citenamefont {Marenduzzo},\ and\
  \citenamefont {Cates}}]{Stenhammar2015}%
  \BibitemOpen
  \bibfield  {author} {\bibinfo {author} {\bibfnamefont {J.}~\bibnamefont
  {Stenhammar}}, \bibinfo {author} {\bibfnamefont {R.}~\bibnamefont
  {Wittkowski}}, \bibinfo {author} {\bibfnamefont {D.}~\bibnamefont
  {Marenduzzo}}, \ and\ \bibinfo {author} {\bibfnamefont {M.~E.}\ \bibnamefont
  {Cates}},\ }\href {\doibase 10.1103/PhysRevLett.114.018301} {\bibfield
  {journal} {\bibinfo  {journal} {Phys. Rev. Lett.}\ }\textbf {\bibinfo
  {volume} {114}},\ \bibinfo {pages} {1} (\bibinfo {year} {2015})}\BibitemShut
  {NoStop}%
\bibitem [{\citenamefont {K{\"{u}}mmel}\ \emph {et~al.}(2015)\citenamefont
  {K{\"{u}}mmel}, \citenamefont {Shabestari}, \citenamefont {Lozano},
  \citenamefont {Volpe},\ and\ \citenamefont {Bechinger}}]{Kummel2015}%
  \BibitemOpen
  \bibfield  {author} {\bibinfo {author} {\bibfnamefont {F.}~\bibnamefont
  {K{\"{u}}mmel}}, \bibinfo {author} {\bibfnamefont {P.}~\bibnamefont
  {Shabestari}}, \bibinfo {author} {\bibfnamefont {C.}~\bibnamefont {Lozano}},
  \bibinfo {author} {\bibfnamefont {G.}~\bibnamefont {Volpe}}, \ and\ \bibinfo
  {author} {\bibfnamefont {C.}~\bibnamefont {Bechinger}},\ }\href {\doibase
  10.1039/C5SM00827A} {\bibfield  {journal} {\bibinfo  {journal} {Soft Matter}\
  }\textbf {\bibinfo {volume} {11}},\ \bibinfo {pages} {1} (\bibinfo {year}
  {2015})}\BibitemShut {NoStop}%
\bibitem [{\citenamefont {Wysocki}, \citenamefont {Winkler},\ and\
  \citenamefont {Gompper}(2016)}]{Wysocki2016}%
  \BibitemOpen
  \bibfield  {author} {\bibinfo {author} {\bibfnamefont {A.}~\bibnamefont
  {Wysocki}}, \bibinfo {author} {\bibfnamefont {R.~G.}\ \bibnamefont
  {Winkler}}, \ and\ \bibinfo {author} {\bibfnamefont {G.}~\bibnamefont
  {Gompper}},\ }\href {\doibase 10.1088/1367-2630/aa529d} {\bibfield  {journal}
  {\bibinfo  {journal} {New J. Phys.}\ }\textbf {\bibinfo {volume} {18}},\
  \bibinfo {pages} {123030} (\bibinfo {year} {2016})}\BibitemShut {NoStop}%
\bibitem [{\citenamefont {Zeitz}, \citenamefont {Wolff},\ and\ \citenamefont
  {Stark}(2017)}]{Zeitz2017}%
  \BibitemOpen
  \bibfield  {author} {\bibinfo {author} {\bibfnamefont {M.}~\bibnamefont
  {Zeitz}}, \bibinfo {author} {\bibfnamefont {K.}~\bibnamefont {Wolff}}, \ and\
  \bibinfo {author} {\bibfnamefont {H.}~\bibnamefont {Stark}},\ }\href
  {\doibase 10.1140/epje/i2017-11510-0} {\bibfield  {journal} {\bibinfo
  {journal} {The European Physical Journal E}\ }\textbf {\bibinfo {volume}
  {40}},\ \bibinfo {pages} {23} (\bibinfo {year} {2017})}\BibitemShut {NoStop}%
\bibitem [{\citenamefont {Morin}\ \emph
  {et~al.}(2017{\natexlab{a}})\citenamefont {Morin}, \citenamefont
  {Lopes~Cardozo}, \citenamefont {Chikkadi},\ and\ \citenamefont
  {Bartolo}}]{morin2017_a}%
  \BibitemOpen
  \bibfield  {author} {\bibinfo {author} {\bibfnamefont {A.}~\bibnamefont
  {Morin}}, \bibinfo {author} {\bibfnamefont {D.}~\bibnamefont
  {Lopes~Cardozo}}, \bibinfo {author} {\bibfnamefont {V.}~\bibnamefont
  {Chikkadi}}, \ and\ \bibinfo {author} {\bibfnamefont {D.}~\bibnamefont
  {Bartolo}},\ }\href {\doibase 10.1103/PhysRevE.96.042611} {\bibfield
  {journal} {\bibinfo  {journal} {Phys. Rev. E}\ }\textbf {\bibinfo {volume}
  {96}},\ \bibinfo {pages} {042611} (\bibinfo {year}
  {2017}{\natexlab{a}})}\BibitemShut {NoStop}%
\bibitem [{\citenamefont {Morin}\ \emph
  {et~al.}(2017{\natexlab{b}})\citenamefont {Morin}, \citenamefont
  {Desreumaux}, \citenamefont {Caussin},\ and\ \citenamefont
  {Bartolo}}]{morin2017_b}%
  \BibitemOpen
  \bibfield  {author} {\bibinfo {author} {\bibfnamefont {A.}~\bibnamefont
  {Morin}}, \bibinfo {author} {\bibfnamefont {N.}~\bibnamefont {Desreumaux}},
  \bibinfo {author} {\bibfnamefont {J.-B.}\ \bibnamefont {Caussin}}, \ and\
  \bibinfo {author} {\bibfnamefont {D.}~\bibnamefont {Bartolo}},\ }\href
  {\doibase 10.1038/nphys3903} {\bibfield  {journal} {\bibinfo  {journal} {Nat.
  Phys.}\ }\textbf {\bibinfo {volume} {13}},\ \bibinfo {pages} {63} (\bibinfo
  {year} {2017}{\natexlab{b}})}\BibitemShut {NoStop}%
\bibitem [{\citenamefont {Marchetti}\ \emph {et~al.}(2013)\citenamefont
  {Marchetti}, \citenamefont {Joanny}, \citenamefont {Ramaswamy}, \citenamefont
  {Liverpool}, \citenamefont {Prost}, \citenamefont {Rao},\ and\ \citenamefont
  {Simha}}]{Marchetti2013}%
  \BibitemOpen
  \bibfield  {author} {\bibinfo {author} {\bibfnamefont {M.~C.}\ \bibnamefont
  {Marchetti}}, \bibinfo {author} {\bibfnamefont {J.~F.}\ \bibnamefont
  {Joanny}}, \bibinfo {author} {\bibfnamefont {S.}~\bibnamefont {Ramaswamy}},
  \bibinfo {author} {\bibfnamefont {T.~B.}\ \bibnamefont {Liverpool}}, \bibinfo
  {author} {\bibfnamefont {J.}~\bibnamefont {Prost}}, \bibinfo {author}
  {\bibfnamefont {M.}~\bibnamefont {Rao}}, \ and\ \bibinfo {author}
  {\bibfnamefont {R.~A.}\ \bibnamefont {Simha}},\ }\href {\doibase
  10.1103/RevModPhys.85.1143} {\bibfield  {journal} {\bibinfo  {journal} {Rev.
  Mod. Phys.}\ }\textbf {\bibinfo {volume} {85}},\ \bibinfo {pages} {1143}
  (\bibinfo {year} {2013})}\BibitemShut {NoStop}%
\bibitem [{\citenamefont {Ramaswamy}(2010)}]{Ramaswany2010}%
  \BibitemOpen
  \bibfield  {author} {\bibinfo {author} {\bibfnamefont {S.}~\bibnamefont
  {Ramaswamy}},\ }\href {\doibase 10.1146/annurev-conmatphys-070909-104101}
  {\bibfield  {journal} {\bibinfo  {journal} {Annu. Rev. Condens. Matter
  Phys.}\ }\textbf {\bibinfo {volume} {1}},\ \bibinfo {pages} {323} (\bibinfo
  {year} {2010})}\BibitemShut {NoStop}%
\bibitem [{\citenamefont {Bechinger}\ \emph {et~al.}(2016)\citenamefont
  {Bechinger}, \citenamefont {Di~Leonardo}, \citenamefont {L\"owen},
  \citenamefont {Reichhardt}, \citenamefont {Volpe},\ and\ \citenamefont
  {Volpe}}]{Bechinger2016}%
  \BibitemOpen
  \bibfield  {author} {\bibinfo {author} {\bibfnamefont {C.}~\bibnamefont
  {Bechinger}}, \bibinfo {author} {\bibfnamefont {R.}~\bibnamefont
  {Di~Leonardo}}, \bibinfo {author} {\bibfnamefont {H.}~\bibnamefont
  {L\"owen}}, \bibinfo {author} {\bibfnamefont {C.}~\bibnamefont {Reichhardt}},
  \bibinfo {author} {\bibfnamefont {G.}~\bibnamefont {Volpe}}, \ and\ \bibinfo
  {author} {\bibfnamefont {G.}~\bibnamefont {Volpe}},\ }\href {\doibase
  10.1103/RevModPhys.88.045006} {\bibfield  {journal} {\bibinfo  {journal}
  {Rev. Mod. Phys.}\ }\textbf {\bibinfo {volume} {88}},\ \bibinfo {pages}
  {045006} (\bibinfo {year} {2016})}\BibitemShut {NoStop}%
\bibitem [{\citenamefont {Briand}, \citenamefont {Schindler},\ and\
  \citenamefont {Dauchot}(2017)}]{briand2017}%
  \BibitemOpen
  \bibfield  {author} {\bibinfo {author} {\bibfnamefont {G.}~\bibnamefont
  {Briand}}, \bibinfo {author} {\bibfnamefont {M.}~\bibnamefont {Schindler}}, \
  and\ \bibinfo {author} {\bibfnamefont {O.}~\bibnamefont {Dauchot}},\
  }\href@noop {} {\bibfield  {journal} {\bibinfo  {journal} {arXiv:1709.03844}\
  } (\bibinfo {year} {2017})}\BibitemShut {NoStop}%
\bibitem [{\citenamefont {Lushi}\ and\ \citenamefont
  {Peskin}(2013)}]{Lushi2013}%
  \BibitemOpen
  \bibfield  {author} {\bibinfo {author} {\bibfnamefont {E.}~\bibnamefont
  {Lushi}}\ and\ \bibinfo {author} {\bibfnamefont {C.~S.}\ \bibnamefont
  {Peskin}},\ }\href {\doibase 10.1016/j.compstruc.2013.03.007} {\bibfield
  {journal} {\bibinfo  {journal} {Comput. Struct.}\ }\textbf {\bibinfo {volume}
  {122}},\ \bibinfo {pages} {239} (\bibinfo {year} {2013})}\BibitemShut
  {NoStop}%
\bibitem [{\citenamefont {Wioland}\ \emph {et~al.}(2013)\citenamefont
  {Wioland}, \citenamefont {Woodhouse}, \citenamefont {Dunkel}, \citenamefont
  {Kessler},\ and\ \citenamefont {Goldstein}}]{Wioland2013}%
  \BibitemOpen
  \bibfield  {author} {\bibinfo {author} {\bibfnamefont {H.}~\bibnamefont
  {Wioland}}, \bibinfo {author} {\bibfnamefont {F.~G.}\ \bibnamefont
  {Woodhouse}}, \bibinfo {author} {\bibfnamefont {J.}~\bibnamefont {Dunkel}},
  \bibinfo {author} {\bibfnamefont {J.~O.}\ \bibnamefont {Kessler}}, \ and\
  \bibinfo {author} {\bibfnamefont {R.~E.}\ \bibnamefont {Goldstein}},\ }\href
  {\doibase 10.1103/PhysRevLett.110.268102} {\bibfield  {journal} {\bibinfo
  {journal} {Phys. Rev. Lett.}\ }\textbf {\bibinfo {volume} {110}},\ \bibinfo
  {pages} {268102} (\bibinfo {year} {2013})}\BibitemShut {NoStop}%
\bibitem [{\citenamefont {Reichhardt}\ and\ \citenamefont
  {Reichhardt}(2017)}]{Reichhardt2017}%
  \BibitemOpen
  \bibfield  {author} {\bibinfo {author} {\bibfnamefont {C.~J.~O.}\
  \bibnamefont {Reichhardt}}\ and\ \bibinfo {author} {\bibfnamefont
  {C.}~\bibnamefont {Reichhardt}},\ }\href {\doibase 10.1088/1367-2630/aaa392}
  {\bibfield  {journal} {\bibinfo  {journal} {New J. Phys.}\ }\textbf {\bibinfo
  {volume} {20}},\ \bibinfo {pages} {025002} (\bibinfo {year}
  {2017})}\BibitemShut {NoStop}%
\bibitem [{\citenamefont {Reichhardt}\ and\ \citenamefont
  {Reichhardt}(2018)}]{Reichhardt2018}%
  \BibitemOpen
  \bibfield  {author} {\bibinfo {author} {\bibfnamefont {C.}~\bibnamefont
  {Reichhardt}}\ and\ \bibinfo {author} {\bibfnamefont {C.~J.~O.}\ \bibnamefont
  {Reichhardt}},\ }\href {\doibase 10.1088/1361-648X/aa9c5f} {\bibfield
  {journal} {\bibinfo  {journal} {J. Phys.: Condens. Matter}\ }\textbf
  {\bibinfo {volume} {30}},\ \bibinfo {pages} {015404} (\bibinfo {year}
  {2018})}\BibitemShut {NoStop}%
\bibitem [{\citenamefont {Kaiser}, \citenamefont {Wensink},\ and\ \citenamefont
  {L\"owen}(2012)}]{Kaiser2012}%
  \BibitemOpen
  \bibfield  {author} {\bibinfo {author} {\bibfnamefont {A.}~\bibnamefont
  {Kaiser}}, \bibinfo {author} {\bibfnamefont {H.~H.}\ \bibnamefont {Wensink}},
  \ and\ \bibinfo {author} {\bibfnamefont {H.}~\bibnamefont {L\"owen}},\ }\href
  {\doibase 10.1103/PhysRevLett.108.268307} {\bibfield  {journal} {\bibinfo
  {journal} {Phys. Rev. Lett.}\ }\textbf {\bibinfo {volume} {108}},\ \bibinfo
  {pages} {268307} (\bibinfo {year} {2012})}\BibitemShut {NoStop}%
\bibitem [{\citenamefont {Kaiser}\ \emph {et~al.}(2013)\citenamefont {Kaiser},
  \citenamefont {Popowa}, \citenamefont {Wensink},\ and\ \citenamefont
  {L\"owen}}]{Kaiser2013}%
  \BibitemOpen
  \bibfield  {author} {\bibinfo {author} {\bibfnamefont {A.}~\bibnamefont
  {Kaiser}}, \bibinfo {author} {\bibfnamefont {K.}~\bibnamefont {Popowa}},
  \bibinfo {author} {\bibfnamefont {H.~H.}\ \bibnamefont {Wensink}}, \ and\
  \bibinfo {author} {\bibfnamefont {H.}~\bibnamefont {L\"owen}},\ }\href
  {\doibase 10.1103/PhysRevE.88.022311} {\bibfield  {journal} {\bibinfo
  {journal} {Phys. Rev. E}\ }\textbf {\bibinfo {volume} {88}},\ \bibinfo
  {pages} {022311} (\bibinfo {year} {2013})}\BibitemShut {NoStop}%
\bibitem [{\citenamefont {McCandlish}, \citenamefont {Baskaran},\ and\
  \citenamefont {Hagan}(2012)}]{McCandlish2012}%
  \BibitemOpen
  \bibfield  {author} {\bibinfo {author} {\bibfnamefont {S.~R.}\ \bibnamefont
  {McCandlish}}, \bibinfo {author} {\bibfnamefont {A.}~\bibnamefont
  {Baskaran}}, \ and\ \bibinfo {author} {\bibfnamefont {M.~F.}\ \bibnamefont
  {Hagan}},\ }\href {\doibase 10.1039/c2sm06960a} {\bibfield  {journal}
  {\bibinfo  {journal} {Soft Matter}\ }\textbf {\bibinfo {volume} {8}},\
  \bibinfo {pages} {2527} (\bibinfo {year} {2012})}\BibitemShut {NoStop}%
\bibitem [{\citenamefont {Menzel}\ and\ \citenamefont
  {L{\"{o}}wen}(2013)}]{Menzel2013}%
  \BibitemOpen
  \bibfield  {author} {\bibinfo {author} {\bibfnamefont {A.~M.}\ \bibnamefont
  {Menzel}}\ and\ \bibinfo {author} {\bibfnamefont {H.}~\bibnamefont
  {L{\"{o}}wen}},\ }\href {\doibase 10.1103/PhysRevLett.110.055702} {\bibfield
  {journal} {\bibinfo  {journal} {Phys. Rev. Lett.}\ }\textbf {\bibinfo
  {volume} {110}},\ \bibinfo {pages} {55702} (\bibinfo {year}
  {2013})}\BibitemShut {NoStop}%
\bibitem [{\citenamefont {Alaimo}, \citenamefont {Praetorius},\ and\
  \citenamefont {Voigt}(2016)}]{Alaimo2016}%
  \BibitemOpen
  \bibfield  {author} {\bibinfo {author} {\bibfnamefont {F.}~\bibnamefont
  {Alaimo}}, \bibinfo {author} {\bibfnamefont {S.}~\bibnamefont {Praetorius}},
  \ and\ \bibinfo {author} {\bibfnamefont {A.}~\bibnamefont {Voigt}},\ }\href
  {\doibase 10.1088/1367-2630/18/8/083008} {\bibfield  {journal} {\bibinfo
  {journal} {New J. Phys.}\ }\textbf {\bibinfo {volume} {18}},\ \bibinfo
  {pages} {083008} (\bibinfo {year} {2016})}\BibitemShut {NoStop}%
\bibitem [{\citenamefont {Ni}\ \emph {et~al.}(2014)\citenamefont {Ni},
  \citenamefont {{Cohen Stuart}}, \citenamefont {Dijkstra},\ and\ \citenamefont
  {Bolhuis}}]{Ni2014}%
  \BibitemOpen
  \bibfield  {author} {\bibinfo {author} {\bibfnamefont {R.}~\bibnamefont
  {Ni}}, \bibinfo {author} {\bibfnamefont {M.~A.}\ \bibnamefont {{Cohen
  Stuart}}}, \bibinfo {author} {\bibfnamefont {M.}~\bibnamefont {Dijkstra}}, \
  and\ \bibinfo {author} {\bibfnamefont {P.~G.}\ \bibnamefont {Bolhuis}},\
  }\href {\doibase 10.1039/C4SM01015A} {\bibfield  {journal} {\bibinfo
  {journal} {Soft Matter}\ }\textbf {\bibinfo {volume} {10}},\ \bibinfo {pages}
  {6609} (\bibinfo {year} {2014})}\BibitemShut {NoStop}%
\bibitem [{\citenamefont {van~der Meer}, \citenamefont {Dijkstra},\ and\
  \citenamefont {Filion}(2016)}]{VanderMeer2016}%
  \BibitemOpen
  \bibfield  {author} {\bibinfo {author} {\bibfnamefont {B.}~\bibnamefont
  {van~der Meer}}, \bibinfo {author} {\bibfnamefont {M.}~\bibnamefont
  {Dijkstra}}, \ and\ \bibinfo {author} {\bibfnamefont {L.}~\bibnamefont
  {Filion}},\ }\href {\doibase 10.1039/C6SM00700G} {\bibfield  {journal}
  {\bibinfo  {journal} {Soft Matter}\ }\textbf {\bibinfo {volume} {12}},\
  \bibinfo {pages} {5630} (\bibinfo {year} {2016})}\BibitemShut {NoStop}%
\bibitem [{\citenamefont {Wu}\ and\ \citenamefont {Libchaber}(2000)}]{Wu2000}%
  \BibitemOpen
  \bibfield  {author} {\bibinfo {author} {\bibfnamefont {X.~L.}\ \bibnamefont
  {Wu}}\ and\ \bibinfo {author} {\bibfnamefont {A.}~\bibnamefont {Libchaber}},\
  }\href {\doibase 10.1103/PhysRevLett.84.3017} {\bibfield  {journal} {\bibinfo
   {journal} {Phys. Rev. Lett.}\ }\textbf {\bibinfo {volume} {84}},\ \bibinfo
  {pages} {3017} (\bibinfo {year} {2000})}\BibitemShut {NoStop}%
\bibitem [{\citenamefont {Valeriani}\ \emph {et~al.}(2011)\citenamefont
  {Valeriani}, \citenamefont {Li}, \citenamefont {Novosel}, \citenamefont
  {Arlt},\ and\ \citenamefont {Marenduzzo}}]{Valeriani2011}%
  \BibitemOpen
  \bibfield  {author} {\bibinfo {author} {\bibfnamefont {C.}~\bibnamefont
  {Valeriani}}, \bibinfo {author} {\bibfnamefont {M.}~\bibnamefont {Li}},
  \bibinfo {author} {\bibfnamefont {J.}~\bibnamefont {Novosel}}, \bibinfo
  {author} {\bibfnamefont {J.}~\bibnamefont {Arlt}}, \ and\ \bibinfo {author}
  {\bibfnamefont {D.}~\bibnamefont {Marenduzzo}},\ }\href {\doibase
  10.1039/c1sm05260h} {\bibfield  {journal} {\bibinfo  {journal} {Soft Matter}\
  }\textbf {\bibinfo {volume} {7}},\ \bibinfo {pages} {5228} (\bibinfo {year}
  {2011})}\BibitemShut {NoStop}%
\bibitem [{\citenamefont {Hinz}\ \emph {et~al.}(2014)\citenamefont {Hinz},
  \citenamefont {Panchenko}, \citenamefont {Kim},\ and\ \citenamefont
  {Fried}}]{Hinz2014}%
  \BibitemOpen
  \bibfield  {author} {\bibinfo {author} {\bibfnamefont {D.~F.}\ \bibnamefont
  {Hinz}}, \bibinfo {author} {\bibfnamefont {A.}~\bibnamefont {Panchenko}},
  \bibinfo {author} {\bibfnamefont {T.-Y.}\ \bibnamefont {Kim}}, \ and\
  \bibinfo {author} {\bibfnamefont {E.}~\bibnamefont {Fried}},\ }\href
  {\doibase 10.1039/c4sm01562b} {\bibfield  {journal} {\bibinfo  {journal}
  {Soft Matter}\ }\textbf {\bibinfo {volume} {10}},\ \bibinfo {pages} {9082}
  (\bibinfo {year} {2014})}\BibitemShut {NoStop}%
\bibitem [{\citenamefont {Swift}\ and\ \citenamefont
  {Hohenberg}(1977)}]{SwiftHohenberg1977}%
  \BibitemOpen
  \bibfield  {author} {\bibinfo {author} {\bibfnamefont {J.}~\bibnamefont
  {Swift}}\ and\ \bibinfo {author} {\bibfnamefont {P.~C.}\ \bibnamefont
  {Hohenberg}},\ }\href {\doibase 10.1103/PhysRevA.15.319} {\bibfield
  {journal} {\bibinfo  {journal} {Phys. Rev. A}\ }\textbf {\bibinfo {volume}
  {15}},\ \bibinfo {pages} {319} (\bibinfo {year} {1977})}\BibitemShut
  {NoStop}%
\bibitem [{\citenamefont {Chan}, \citenamefont {Goldenfeld},\ and\
  \citenamefont {Dantzig}(2009)}]{Chan2009}%
  \BibitemOpen
  \bibfield  {author} {\bibinfo {author} {\bibfnamefont {P.}~\bibnamefont
  {Chan}}, \bibinfo {author} {\bibfnamefont {N.}~\bibnamefont {Goldenfeld}}, \
  and\ \bibinfo {author} {\bibfnamefont {J.}~\bibnamefont {Dantzig}},\ }\href
  {\doibase 10.1103/PhysRevE.79.035701} {\bibfield  {journal} {\bibinfo
  {journal} {Phys. Rev. E}\ }\textbf {\bibinfo {volume} {79}},\ \bibinfo
  {pages} {035701} (\bibinfo {year} {2009})}\BibitemShut {NoStop}%
\bibitem [{\citenamefont {Berry}\ and\ \citenamefont
  {Grant}(2011)}]{Berry2011}%
  \BibitemOpen
  \bibfield  {author} {\bibinfo {author} {\bibfnamefont {J.}~\bibnamefont
  {Berry}}\ and\ \bibinfo {author} {\bibfnamefont {M.}~\bibnamefont {Grant}},\
  }\href {\doibase 10.1103/PhysRevLett.106.175702} {\bibfield  {journal}
  {\bibinfo  {journal} {Phys. Rev. Lett.}\ }\textbf {\bibinfo {volume} {106}},\
  \bibinfo {pages} {175702} (\bibinfo {year} {2011})}\BibitemShut {NoStop}%
\bibitem [{\citenamefont {Robbins}\ \emph {et~al.}(2012)\citenamefont
  {Robbins}, \citenamefont {Archer}, \citenamefont {Thiele},\ and\
  \citenamefont {Knobloch}}]{Robbins2012}%
  \BibitemOpen
  \bibfield  {author} {\bibinfo {author} {\bibfnamefont {M.~J.}\ \bibnamefont
  {Robbins}}, \bibinfo {author} {\bibfnamefont {a.~J.}\ \bibnamefont {Archer}},
  \bibinfo {author} {\bibfnamefont {U.}~\bibnamefont {Thiele}}, \ and\ \bibinfo
  {author} {\bibfnamefont {E.}~\bibnamefont {Knobloch}},\ }\href {\doibase
  10.1103/PhysRevE.85.061408} {\bibfield  {journal} {\bibinfo  {journal} {Phys.
  Rev. E}\ }\textbf {\bibinfo {volume} {85}},\ \bibinfo {pages} {061408}
  (\bibinfo {year} {2012})}\BibitemShut {NoStop}%
\bibitem [{\citenamefont {Praetorius}\ and\ \citenamefont
  {Voigt}(2015{\natexlab{a}})}]{Praetorius2015}%
  \BibitemOpen
  \bibfield  {author} {\bibinfo {author} {\bibfnamefont {S.}~\bibnamefont
  {Praetorius}}\ and\ \bibinfo {author} {\bibfnamefont {A.}~\bibnamefont
  {Voigt}},\ }\href {\doibase 10.1063/1.4918559} {\bibfield  {journal}
  {\bibinfo  {journal} {J. Chem. Phys.}\ }\textbf {\bibinfo {volume} {142}},\
  \bibinfo {pages} {154904} (\bibinfo {year} {2015}{\natexlab{a}})}\BibitemShut
  {NoStop}%
\bibitem [{\citenamefont {Menzel}, \citenamefont {Ohta},\ and\ \citenamefont
  {L\"{o}wen}(2014)}]{Menzel2014}%
  \BibitemOpen
  \bibfield  {author} {\bibinfo {author} {\bibfnamefont {A.~M.}\ \bibnamefont
  {Menzel}}, \bibinfo {author} {\bibfnamefont {T.}~\bibnamefont {Ohta}}, \ and\
  \bibinfo {author} {\bibfnamefont {H.}~\bibnamefont {L\"{o}wen}},\ }\href
  {\doibase 10.1103/PhysRevE.89.022301} {\bibfield  {journal} {\bibinfo
  {journal} {Phys. Rev. E}\ }\textbf {\bibinfo {volume} {89}},\ \bibinfo
  {pages} {022301} (\bibinfo {year} {2014})}\BibitemShut {NoStop}%
\bibitem [{\citenamefont {Elder}\ \emph {et~al.}(2002)\citenamefont {Elder},
  \citenamefont {Katakowski}, \citenamefont {Haataja},\ and\ \citenamefont
  {Grant}}]{Elder2002}%
  \BibitemOpen
  \bibfield  {author} {\bibinfo {author} {\bibfnamefont {K.~R.}\ \bibnamefont
  {Elder}}, \bibinfo {author} {\bibfnamefont {M.}~\bibnamefont {Katakowski}},
  \bibinfo {author} {\bibfnamefont {M.}~\bibnamefont {Haataja}}, \ and\
  \bibinfo {author} {\bibfnamefont {M.}~\bibnamefont {Grant}},\ }\href
  {\doibase 10.1103/PhysRevLett.88.245701} {\bibfield  {journal} {\bibinfo
  {journal} {Phys. Rev. Lett.}\ }\textbf {\bibinfo {volume} {88}},\ \bibinfo
  {pages} {245701} (\bibinfo {year} {2002})}\BibitemShut {NoStop}%
\bibitem [{\citenamefont {Elder}\ and\ \citenamefont
  {Grant}(2004)}]{Elder2004}%
  \BibitemOpen
  \bibfield  {author} {\bibinfo {author} {\bibfnamefont {K.~R.}\ \bibnamefont
  {Elder}}\ and\ \bibinfo {author} {\bibfnamefont {M.}~\bibnamefont {Grant}},\
  }\href {\doibase 10.1103/PhysRevE.70.051605} {\bibfield  {journal} {\bibinfo
  {journal} {Phys. Rev. E}\ }\textbf {\bibinfo {volume} {70}},\ \bibinfo
  {pages} {051605} (\bibinfo {year} {2004})}\BibitemShut {NoStop}%
\bibitem [{\citenamefont {Elder}\ \emph {et~al.}(2007)\citenamefont {Elder},
  \citenamefont {Provatas}, \citenamefont {Berry}, \citenamefont {Stefanovic},\
  and\ \citenamefont {Grant}}]{Elder2007}%
  \BibitemOpen
  \bibfield  {author} {\bibinfo {author} {\bibfnamefont {K.~R.}\ \bibnamefont
  {Elder}}, \bibinfo {author} {\bibfnamefont {N.}~\bibnamefont {Provatas}},
  \bibinfo {author} {\bibfnamefont {J.}~\bibnamefont {Berry}}, \bibinfo
  {author} {\bibfnamefont {P.}~\bibnamefont {Stefanovic}}, \ and\ \bibinfo
  {author} {\bibfnamefont {M.}~\bibnamefont {Grant}},\ }\href {\doibase
  10.1103/PhysRevB.75.064107} {\bibfield  {journal} {\bibinfo  {journal} {Phys.
  Rev. B}\ }\textbf {\bibinfo {volume} {75}},\ \bibinfo {pages} {64107}
  (\bibinfo {year} {2007})}\BibitemShut {NoStop}%
\bibitem [{\citenamefont {van Teeffelen}\ \emph {et~al.}(2009)\citenamefont
  {van Teeffelen}, \citenamefont {Backofen}, \citenamefont {Voigt},\ and\
  \citenamefont {L\"owen}}]{vanTeeffelen2009}%
  \BibitemOpen
  \bibfield  {author} {\bibinfo {author} {\bibfnamefont {S.}~\bibnamefont {van
  Teeffelen}}, \bibinfo {author} {\bibfnamefont {R.}~\bibnamefont {Backofen}},
  \bibinfo {author} {\bibfnamefont {A.}~\bibnamefont {Voigt}}, \ and\ \bibinfo
  {author} {\bibfnamefont {H.}~\bibnamefont {L\"owen}},\ }\href {\doibase
  10.1103/PhysRevE.79.051404} {\bibfield  {journal} {\bibinfo  {journal} {Phys.
  Rev. E}\ }\textbf {\bibinfo {volume} {79}},\ \bibinfo {pages} {051404}
  (\bibinfo {year} {2009})}\BibitemShut {NoStop}%
\bibitem [{\citenamefont {Backofen}, \citenamefont {R\"atz},\ and\
  \citenamefont {Voigt}(2007)}]{Backofen2007}%
  \BibitemOpen
  \bibfield  {author} {\bibinfo {author} {\bibfnamefont {R.}~\bibnamefont
  {Backofen}}, \bibinfo {author} {\bibfnamefont {A.}~\bibnamefont {R\"atz}}, \
  and\ \bibinfo {author} {\bibfnamefont {A.}~\bibnamefont {Voigt}},\
  }\href@noop {} {\bibfield  {journal} {\bibinfo  {journal} {Phil. Mag. Lett.}\
  }\textbf {\bibinfo {volume} {87}},\ \bibinfo {pages} {813} (\bibinfo {year}
  {2007})}\BibitemShut {NoStop}%
\bibitem [{\citenamefont {Praetorius}\ and\ \citenamefont
  {Voigt}(2015{\natexlab{b}})}]{Praetorius2015_SIAM}%
  \BibitemOpen
  \bibfield  {author} {\bibinfo {author} {\bibfnamefont {S.}~\bibnamefont
  {Praetorius}}\ and\ \bibinfo {author} {\bibfnamefont {A.}~\bibnamefont
  {Voigt}},\ }\href {\doibase 10.1137/140980375} {\bibfield  {journal}
  {\bibinfo  {journal} {SIAM Journal on Scientific Computing}\ }\textbf
  {\bibinfo {volume} {37}},\ \bibinfo {pages} {B425} (\bibinfo {year}
  {2015}{\natexlab{b}})}\BibitemShut {NoStop}%
\bibitem [{\citenamefont {Vey}\ and\ \citenamefont {Voigt}(2007)}]{Vey2007}%
  \BibitemOpen
  \bibfield  {author} {\bibinfo {author} {\bibfnamefont {S.}~\bibnamefont
  {Vey}}\ and\ \bibinfo {author} {\bibfnamefont {A.}~\bibnamefont {Voigt}},\
  }\href {\doibase 10.1007/s00791-006-0048-3} {\bibfield  {journal} {\bibinfo
  {journal} {Comput. Visualization Sci.}\ }\textbf {\bibinfo {volume} {10}},\
  \bibinfo {pages} {57} (\bibinfo {year} {2007})}\BibitemShut {NoStop}%
\bibitem [{\citenamefont {Witkowski}\ \emph {et~al.}(2015)\citenamefont
  {Witkowski}, \citenamefont {Ling}, \citenamefont {Praetorius},\ and\
  \citenamefont {Voigt}}]{Witkowski2015}%
  \BibitemOpen
  \bibfield  {author} {\bibinfo {author} {\bibfnamefont {T.}~\bibnamefont
  {Witkowski}}, \bibinfo {author} {\bibfnamefont {S.}~\bibnamefont {Ling}},
  \bibinfo {author} {\bibfnamefont {S.}~\bibnamefont {Praetorius}}, \ and\
  \bibinfo {author} {\bibfnamefont {A.}~\bibnamefont {Voigt}},\ }\href
  {\doibase 10.1007/s10444-015-9405-4} {\bibfield  {journal} {\bibinfo
  {journal} {Advances in Computational Mathematics}\ }\textbf {\bibinfo
  {volume} {41}},\ \bibinfo {pages} {1145} (\bibinfo {year}
  {2015})}\BibitemShut {NoStop}%
\bibitem [{\citenamefont {Skoge}\ \emph {et~al.}(2006)\citenamefont {Skoge},
  \citenamefont {Donev}, \citenamefont {Stillinger},\ and\ \citenamefont
  {Torquato}}]{Skoge2006}%
  \BibitemOpen
  \bibfield  {author} {\bibinfo {author} {\bibfnamefont {M.}~\bibnamefont
  {Skoge}}, \bibinfo {author} {\bibfnamefont {A.}~\bibnamefont {Donev}},
  \bibinfo {author} {\bibfnamefont {F.~H.}\ \bibnamefont {Stillinger}}, \ and\
  \bibinfo {author} {\bibfnamefont {S.}~\bibnamefont {Torquato}},\ }\href
  {\doibase 10.1103/PhysRevE.74.041127} {\bibfield  {journal} {\bibinfo
  {journal} {Phys. Rev. E}\ }\textbf {\bibinfo {volume} {74}},\ \bibinfo
  {pages} {041127} (\bibinfo {year} {2006})}\BibitemShut {NoStop}%
\bibitem [{\citenamefont {Grossman}, \citenamefont {Aranson},\ and\
  \citenamefont {{Ben Jacob}}(2008)}]{Grossman2008}%
  \BibitemOpen
  \bibfield  {author} {\bibinfo {author} {\bibfnamefont {D.}~\bibnamefont
  {Grossman}}, \bibinfo {author} {\bibfnamefont {I.~S.}\ \bibnamefont
  {Aranson}}, \ and\ \bibinfo {author} {\bibfnamefont {E.}~\bibnamefont {{Ben
  Jacob}}},\ }\href {\doibase 10.1088/1367-2630/10/2/023036} {\bibfield
  {journal} {\bibinfo  {journal} {New J. Phys.}\ }\textbf {\bibinfo {volume}
  {10}},\ \bibinfo {pages} {023036} (\bibinfo {year} {2008})}\BibitemShut
  {NoStop}%
\bibitem [{\citenamefont {L\"{o}ber}, \citenamefont {Ziebert},\ and\
  \citenamefont {Aranson}(2015)}]{Lober2015}%
  \BibitemOpen
  \bibfield  {author} {\bibinfo {author} {\bibfnamefont {J.}~\bibnamefont
  {L\"{o}ber}}, \bibinfo {author} {\bibfnamefont {F.}~\bibnamefont {Ziebert}},
  \ and\ \bibinfo {author} {\bibfnamefont {I.~S.}\ \bibnamefont {Aranson}},\
  }\href {\doibase 10.1038/srep09172} {\bibfield  {journal} {\bibinfo
  {journal} {Scientific Reports}\ }\textbf {\bibinfo {volume} {5}},\ \bibinfo
  {pages} {9172} (\bibinfo {year} {2015})}\BibitemShut {NoStop}%
\bibitem [{\citenamefont {Marth}\ and\ \citenamefont
  {Voigt}(2016)}]{Marth2016}%
  \BibitemOpen
  \bibfield  {author} {\bibinfo {author} {\bibfnamefont {W.}~\bibnamefont
  {Marth}}\ and\ \bibinfo {author} {\bibfnamefont {A.}~\bibnamefont {Voigt}},\
  }\href {\doibase 10.1098/rsfs.2016.0037} {\bibfield  {journal} {\bibinfo
  {journal} {Interface Focus}\ }\textbf {\bibinfo {volume} {6}},\ \bibinfo
  {pages} {20160037} (\bibinfo {year} {2016})}\BibitemShut {NoStop}%
\bibitem [{\citenamefont {Heinonen}\ \emph {et~al.}(2016)\citenamefont
  {Heinonen}, \citenamefont {Achim}, \citenamefont {Kosterlitz}, \citenamefont
  {Ying}, \citenamefont {Lowengrub},\ and\ \citenamefont
  {Ala-Nissila}}]{Heinonen2016}%
  \BibitemOpen
  \bibfield  {author} {\bibinfo {author} {\bibfnamefont {V.}~\bibnamefont
  {Heinonen}}, \bibinfo {author} {\bibfnamefont {C.~V.}\ \bibnamefont {Achim}},
  \bibinfo {author} {\bibfnamefont {J.~M.}\ \bibnamefont {Kosterlitz}},
  \bibinfo {author} {\bibfnamefont {S.-C.}\ \bibnamefont {Ying}}, \bibinfo
  {author} {\bibfnamefont {J.}~\bibnamefont {Lowengrub}}, \ and\ \bibinfo
  {author} {\bibfnamefont {T.}~\bibnamefont {Ala-Nissila}},\ }\href {\doibase
  10.1103/PhysRevLett.116.024303} {\bibfield  {journal} {\bibinfo  {journal}
  {Phys. Rev. Lett.}\ }\textbf {\bibinfo {volume} {116}},\ \bibinfo {pages}
  {024303} (\bibinfo {year} {2016})}\BibitemShut {NoStop}%
\end{thebibliography}%

\end{document}